\documentclass[12pt]{article}
\usepackage{epsfig}

\def\beq{\begin{equation}}
\def\eeq#1{\label{#1}\end{equation}}
\def\eeqn{\end{equation}}

\def\beqa{\begin{eqnarray}}
\def\eeqa#1{\label{#1}\end{eqnarray}}
\def\eeqan{\end{eqnarray}}

\let\bar=\overbar

\def\Dslash{\not{\hbox{\kern-4pt $D$}}}
\def\dslash{\not{\hbox{\kern-2pt $\del$}}}

\def\msb{{\bar{\ssstyle M \kern -1pt S}}}

\def\Title#1{\begin{center} {\Large {\bf #1} } \end{center}}

\begin{document}

\Title{$|V_{cd}|$ and $|V_{cs}|$ from Semileptonic and Leptonic decays using lattice QCD methods}

\bigskip
\bigskip


\footnotesize \noindent Proceedings of CKM 2012, the 7th International Workshop on the CKM Unitarity Triangle, University of Cincinnati, USA, 28 September - 2 October 2012 \normalsize
\bigskip\bigskip

\begin{raggedright}  

{\it Heechang Na\index{Na, H.}\\
Argonne Leadership Computing Facility\\
Argonne National Laboratory\\
Argonne, IL 60439 USA}
\end{raggedright}

\section{Introduction}

The CKM matrix elements $|V_{cd}|$ and $|V_{cs}|$ can be determined from semileptonic and leptonic decays of $D$ or $D_s$ mesons combining experiments and lattice simulations. 
One can use $D \rightarrow \pi l \nu$ and $D \rightarrow l \nu$ channels to determine $|V_{cd}|$, and  $D \rightarrow K l \nu$ and $D_s \rightarrow l \nu$ channels for $|V_{cs}|$.  
We have relations between experiments and theory for the CKM matrix elements,
\begin{equation}
\frac{d}{dq^2}\Gamma \propto |V_{cx}|^2 |f_+(q^2)|^2, \;\;\;\;\;\;\; \Gamma \propto |V_{cx}|^2 f_{D_x}^2
\end{equation}
for semileptonic (left) and leptonic (right) decays,
where $\Gamma$ is the decay rate measured from experiments.
Thus, once one determines the form factor $f_+(q^2)$ or the decay constant $f_{D_x}$ from theory, one can extract the CKM matrix elements from the relations between experiments and theory.  
Lattice QCD can provide the most reliable calculations for the form factors and decay constants, since the theoretical calculations should address non-perturbative weak matrix element contributions and lattice QCD is the only successful non-perturbative method. 
In this paper, we review recent determinations of $|V_{cd}|$ and $|V_{cs}|$ using lattice QCD methods. 

\section{Semileptonic decays: $D \rightarrow \pi l \nu$ and $D \rightarrow K l \nu$}

The form factors of the D meson semileptonic decays have been calculated by many lattice groups.
Here, we would like to focus on recent works from Fermilab/MILC, HPQCD, and ETM collaborations with dynamical simulations. Fermilab/MILC (2005)~\cite{milc2005} used MILC asqtad $N_f=2+1$ gauge configurations with asqtad light quarks and Fermilab charm quarks. Their final results show 10 \% total errors including all systematic errors. The biggest error is due to the charm quark operator matching. Their results showed the form factor shape dependence on $q^2$ before the measurements of the shape by Belle, BaBar, and CLEO-c. The most recent published results are from HPQCD (2011)~\cite{hpqcd2011}. They used MILC asqtad $N_f=2+1$ lattices, just like Fermilab/MILC, but applied the HISQ action for the valence light and charm quarks.
HPQCD found that they can use a scalar current rather than a vector current, so that the calculation could be done without operator matching. Due to the small discretization errors of the HISQ action and the trick with the scalar current, HPQCD achieved very good accuracy; 2.5 \% total errors for $D$ to $K$ and 4.4 \% total errors for $D$ to $\pi$ semileptonic decays. In addition to those listed above, ETMC~\cite{etmc2010}, Fermilab/MILC~\cite{milc2012}, and HPQCD~\cite{hpqcd2012} also have preliminary results. In particular, HPQCD's preliminary results~\cite{hpqcd2012} are interesting. They integrate over the experimental $q^2$ bins, so they can fit through the entire $q^2$ region. As a result, they obtain more than a factor of two smaller error than their previous calculation~\cite{hpqcd2011}. All their results show very good agreement.

\begin{figure}[htb]
\begin{center}
\includegraphics[width=.56\textwidth,angle=270]{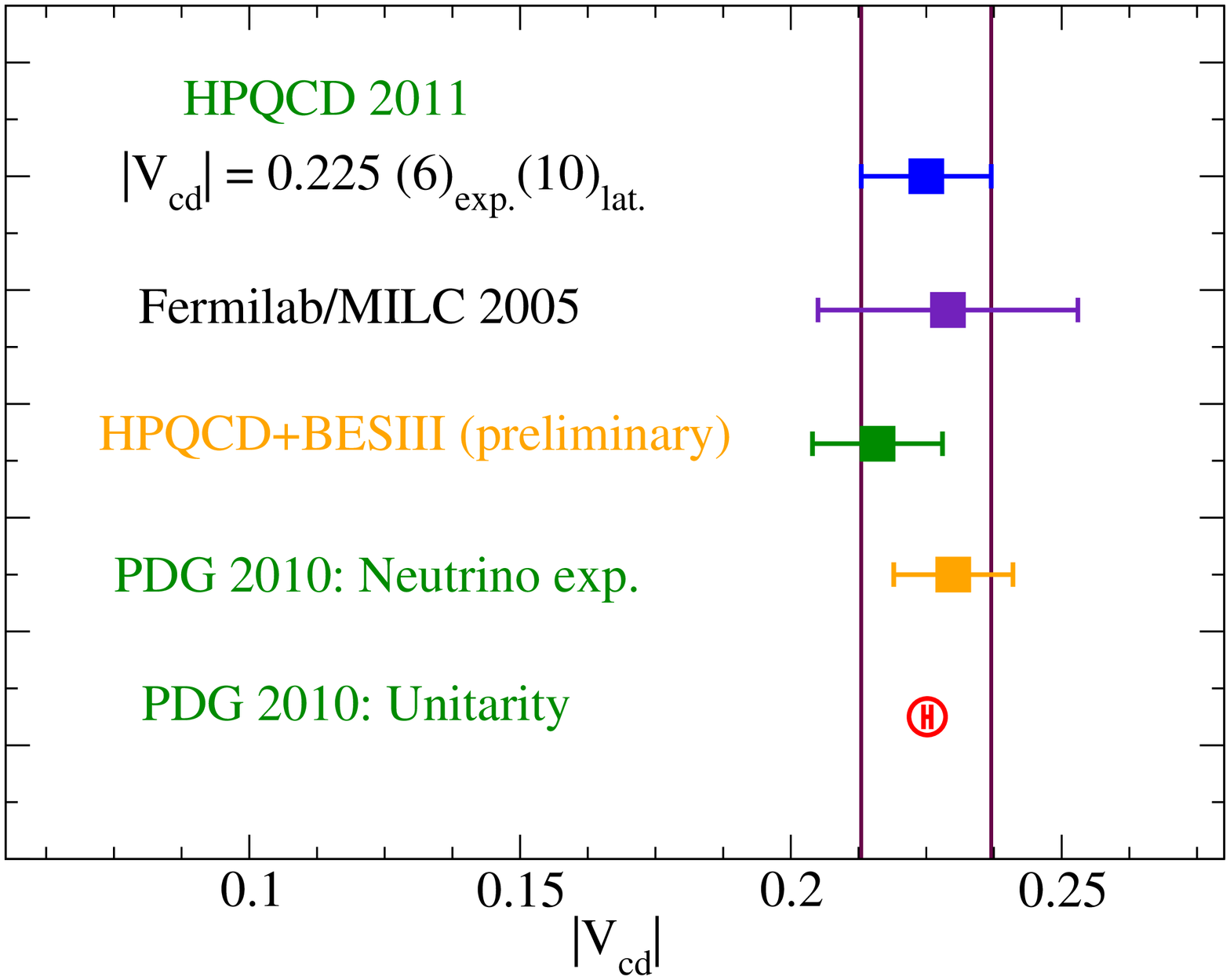}
\includegraphics[width=.56\textwidth,angle=270]{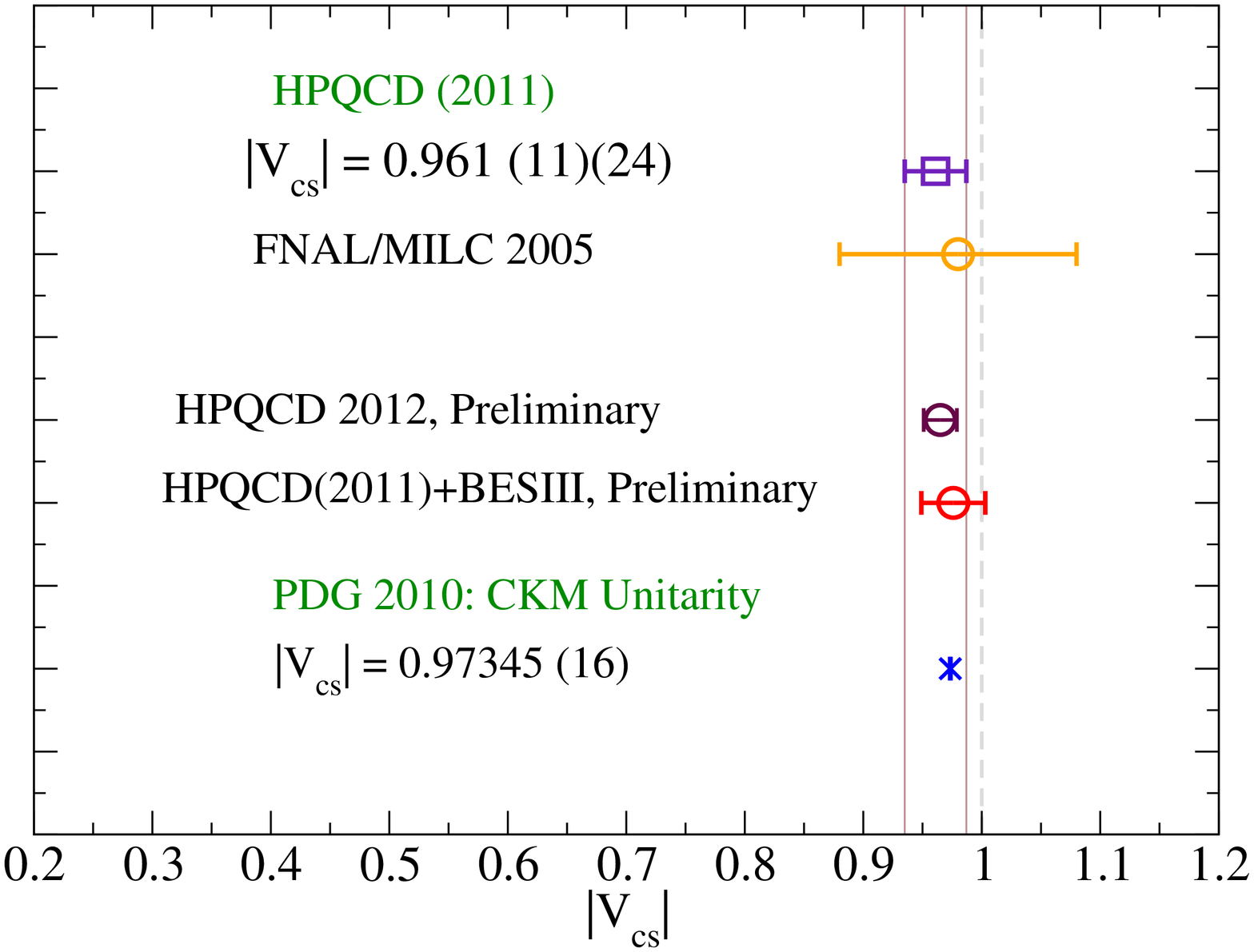}
\caption{Comparisons of results for $|V_{cd}|$ (upper) and $|V_{cs}|$ (lower) from semileptonic decays.}
\label{vv}
\end{center}
\end{figure}

Drawing upon all the experiments available today, we obtain and compare $|V_{cd}|$ and $|V_{cs}|$ in Fig.~\ref{vv}.
One can easily notice that the results for both $|V_{cd}|$ and $|V_{cs}|$  show very good agreement between different lattice calculations, different experiments including the new preliminary result from BES III~\cite{besIII}, and the unitarity point. 
We do not find any hint of New Physics here; however, these investigations present highly non-trivial checks between different experiments and theory calculations. 
We note that, for $|V_{cd}|$, HPQCD's 2011 result~\cite{hpqcd2011} is the first lattice calculation with comparable errors to the result from the neutrino experiment.  
So far, the PDG quotes the neutrino experiment results for $|V_{cd}|$. This is simply because lattice calculations had much larger errors in the past. 

\section{Leptonic decays: $D \rightarrow l \nu$ and $D_s \rightarrow l \nu$.}

\begin{figure}[htb]
\begin{center}
\includegraphics[width=.56\textwidth,angle=270]{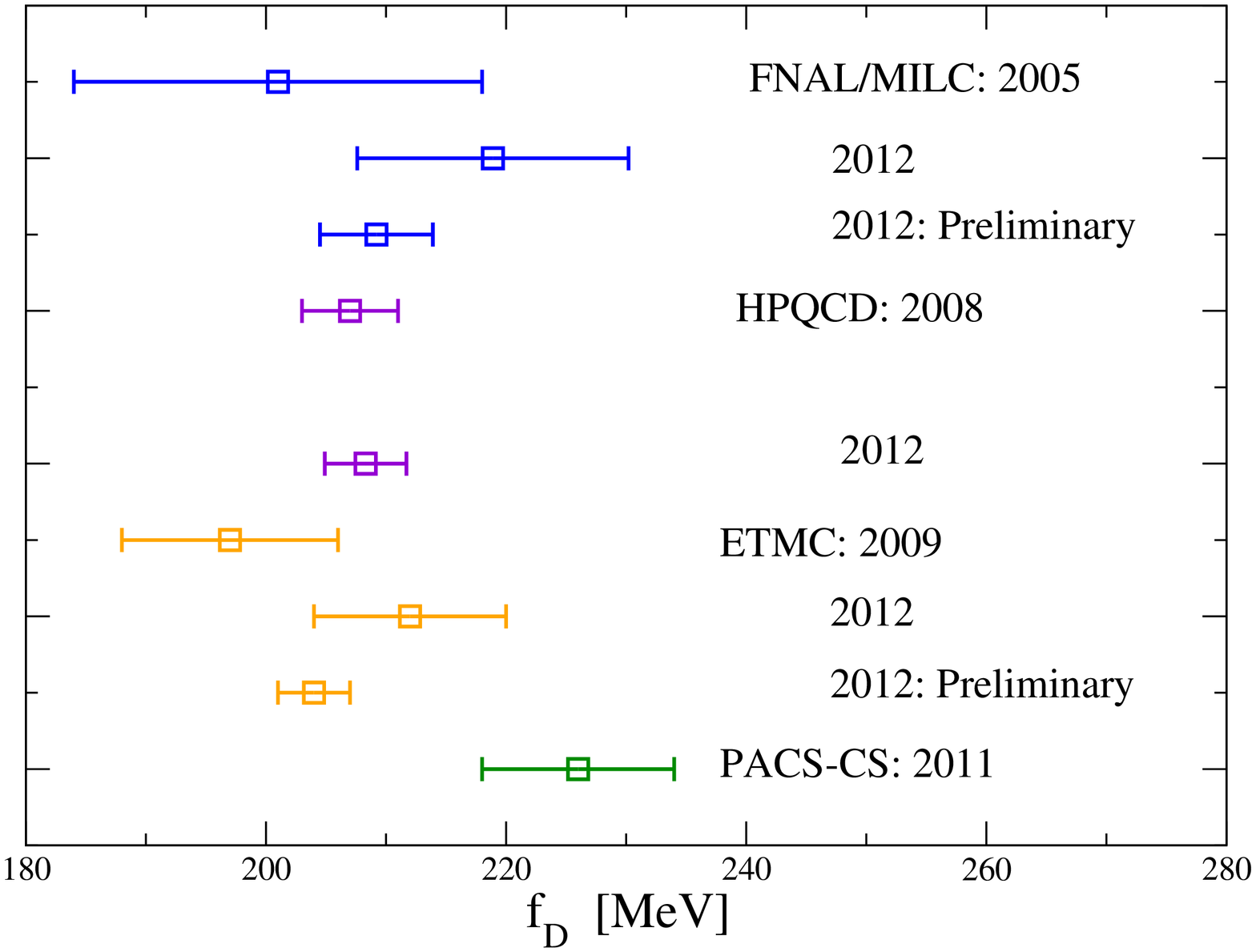}
\includegraphics[width=.56\textwidth,angle=270]{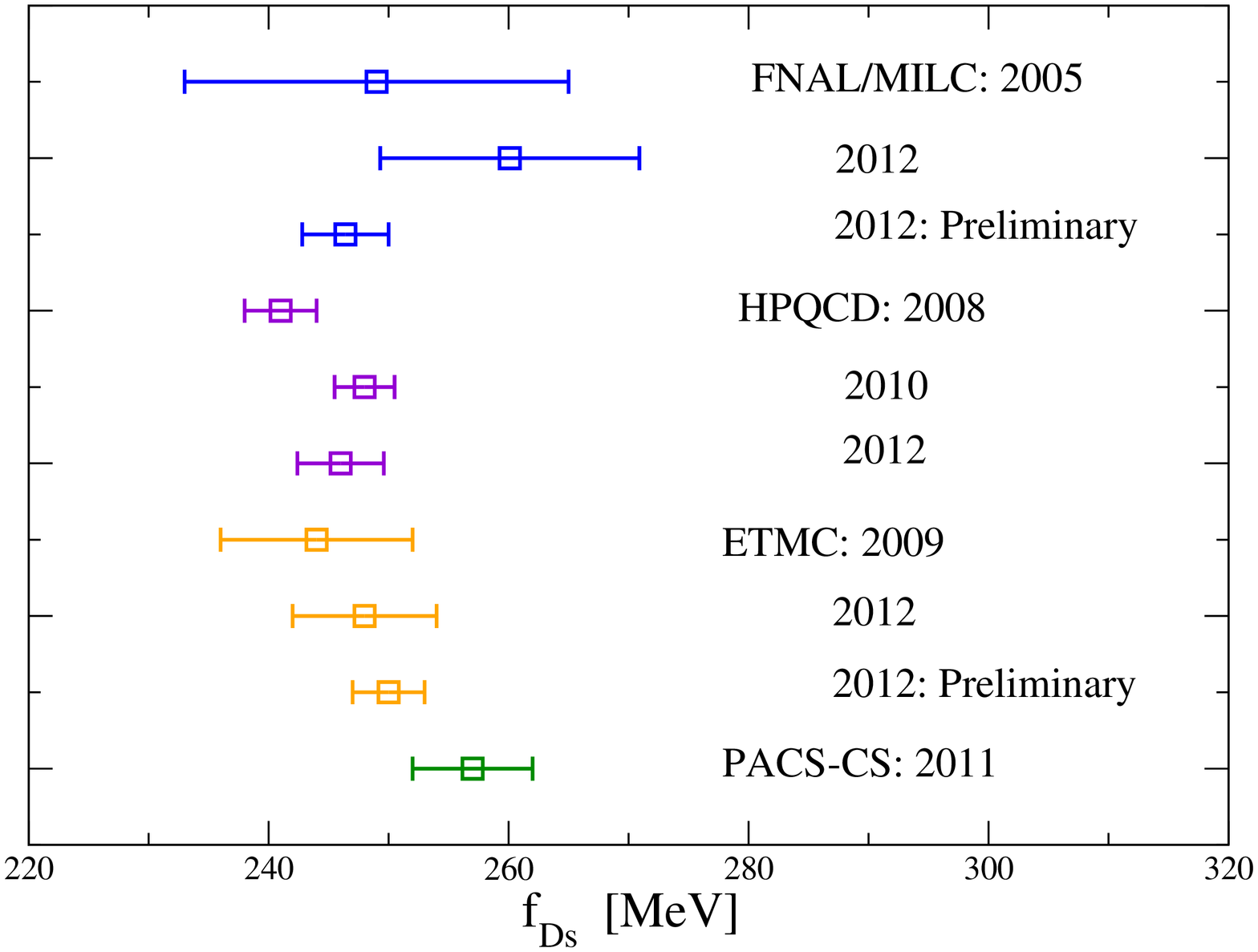}
\caption{Comparisons of results for $f_D$ (upper) and $f_{D_s}$ (lower) from Fermilab/MILC~\cite{milc}, HPQCD~\cite{hpqcd}, ETMC~\cite{etmc}, and PACS-CS~\cite{pacs}.}
\label{ff}
\end{center}
\end{figure}

We have many lattice calculation results for the decay constants $f_D$ and $f_{D_s}$. Fig.~\ref{ff} shows comparisons of the decay constants from various collaborations including very recent preliminary results. The figure contains quite important progress in lattice QCD methods. 
Fermilab/MILC's 2012 preliminary results and PACS-CS's 2011 results were simulated at the physical pion mass, which allows one to extract the decay constants with very small or no chiral extrapolation errors.
The chiral extrapolation is one of the very difficult tasks in lattice QCD calculations in general. 
In addition, $N_f=2+1+1$ lattice configurations were used for Fermilab/MILC's 2012 and ETMC's 2012 preliminary results. 
If we consider only the most accurate results from each collaboration, the size of errors are comparable and the results are in good agreement: comparing Fermilab/MILC 2012, HPQCD 2012, and ETMC 2012 results for $f_D$; and Fermilab/MILC 2012, HPQCD 2010, and ETMC 2012 for $f_{D_s}$. 
We take averages of the three results, and will use the averages for extracting $|V_{cd}|$ and $|V_{cs}|$ in the following. 

\begin{figure}[htb]
\begin{center}
\includegraphics[width=.56\textwidth,angle=270]{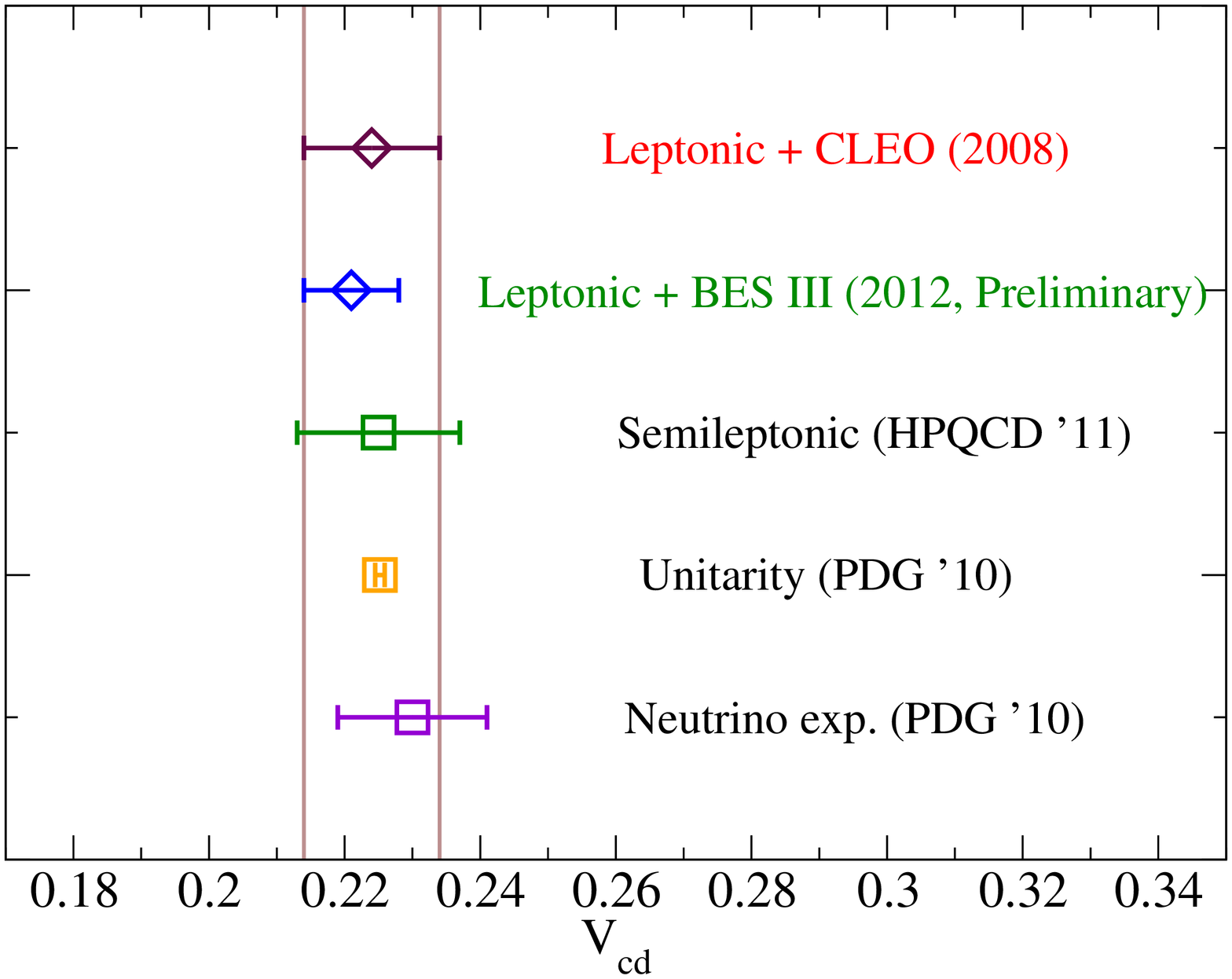}
\includegraphics[width=.56\textwidth,angle=270]{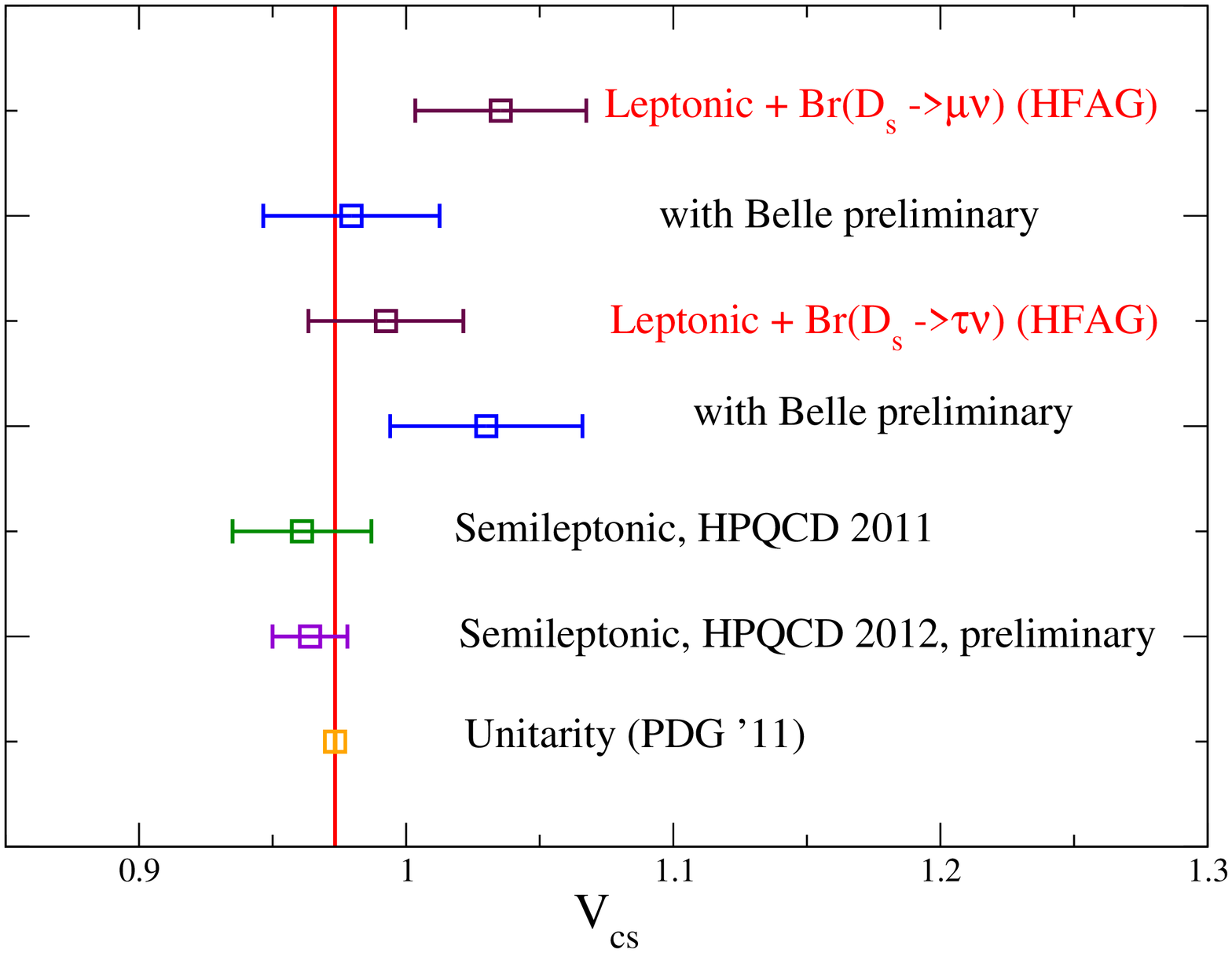}
\caption{Comparisons of results for $|V_{cd}|$ (upper) and $|V_{cs}|$ (lower) from leptonic decays. We also display the most accurate results from semileptonic decays for comparisons.}
\label{vv2}
\end{center}
\end{figure}

Now, we have $|V_{cd}|$ and $|V_{cs}|$ results in Fig.~\ref{vv2}.
Let's look at the upper plot for $|V_{cd}|$. As one can see, all determinations of $|V_{cd}|$ agree very well with the unitarity point. It is even true with the new BES III preliminary result~\cite{besIII2} or from semileptonic decays. We note that the result with leptonic decays and the BES III preliminary experiment~\cite{besIII2} (the second result from the top in the upper plot) give significantly smaller errors than the error from the neutrino experiment. 

For $|V_{cs}|$ on the lower plot of Fig.~\ref{vv2}, the situation is a little different. 
As we saw in Fig.~\ref{vv}, the results from semileptonic decays show good agreement with the unitarity point. However, for the results from leptonic decays, it depends on the experiments.
In the experiments, we have two channels $D_s \rightarrow \mu \nu$ and $D_s \rightarrow \tau \nu$ that we can use to extract $|V_{cs}|$. For the two channels, we have averaged numbers from HFAG, which are labeled in red on the plot, and this year we have preliminary results from Belle.
As one can see, the $D_s \rightarrow \mu \nu$ Belle preliminary~\cite{belle} and  $D_s \rightarrow \tau \nu$ HFAG results are in good agreement to the unitarity point, while the other two cases show some deviations which are a little more than 1 $\sigma$.

We will not try here to get an average between the HFAG and the Belle preliminary results, since the correlations between the two would be complicated. 
However, it is certain that $|V_{cs}|$ results from leptonic decays are not in very good agreement with the unitarity point.  
This is actually identical to the observation of the $f_{D_s}$ puzzle~\cite{puzzle}.
The puzzle represents the difference between $f_{D_s}$ determinations from the lattice and experiment, and in the experiment one needs to use the unitarity $|V_{cs}|$.  
Current deviation is about 1.5 $\sigma$; however, if the errors on the leptonic determinations of $|V_{cs}|$ are reduced by a factor of two while the middle value stays as the current number, then the deviation would be significant. So, we need to wait to see the experiments attain more accuracy, and this could lead to a hint for New Physics.


\end{document}